\documentclass[aps,prl,floatfix,twocolumn,groupedaddress]{revtex4}
\usepackage{graphicx}
\usepackage{times}
\usepackage{amssymb,amsmath}

\newcommand{\ket}[1]{\left\vert#1\right\rangle}
\newcommand{\bra}[1]{\langle #1\vert}

\newcommand{\nm}{\,\mbox{nm}}

\newcommand{\mhz}{\,\mbox{MHz}}
\newcommand{\khz}{\,\mbox{kHz}}

\newcommand{\fs}{\,\mbox{fs}}

\begin{document}
\title{Quantum simulation of the wavefunction to probe frustrated Heisenberg spin systems}

\author{Xiao-song Ma$^{1,2,\dagger}$, Borivoje Dakic$^{1,2, \dagger}$, William Naylor$^{1,2}$, Anton Zeilinger$^{1,2,3}$, Philip Walther$^{1,2}$}
\affiliation{
$^1$~Institute for Quantum Optics and Quantum Information (IQOQI), Austrian Academy of Sciences, Boltzmanngasse 3, A-1090 Vienna, Austria\\
$^2$~Faculty of Physics, University of Vienna, Boltzmanngasse 5, A-1090 Vienna, Austria\\
$^3$~Vienna Centre for Quantum Science and Technology, Boltzmanngasse 3, A-1090 Vienna, Austria\\
$^{\dagger}$~These authors contributed equally to this work}

\begin{abstract}
Quantum simulators are controllable quantum systems that can reproduce the dynamics of the system of interest, which are unfeasible for classical computers. Recent developments in quantum technology enable the precise control of individual quantum particles as required for studying complex quantum systems. Particularly, quantum simulators capable of simulating frustrated Heisenberg spin systems provide platforms for understanding exotic matter such as high-temperature superconductors. Here we report the analog quantum simulation of the ground-state wavefunction to probe arbitrary Heisenberg-type interactions among four spin-1/2 particles . Depending on the interaction strength, frustration within the system emerges such that the ground state evolves from a localized to a resonating valence-bond state. This spin-1/2 tetramer is created using the polarization states of four photons. The single-particle addressability and tunable measurement-induced interactions provide us insights into entanglement dynamics among individual particles. We directly extract ground-state energies and pair-wise quantum correlations to observe the monogamy of entanglement.
\end{abstract}

\maketitle

During the past years, there has been an explosion of interest
in quantum-enhanced technologies. The applications are many-fold and reach from
quantum metrology\cite{Giovannetti2004} to quantum information processing\cite{Zoller2005}. In particular quantum computation has generated
a lot of interest due to the discovery of quantum algorithms\cite{Deutsch1992, Shor1994, Grover1997} which
outperform classical ones. The first
proposed application for which quantum computation can give an exponential
enhancement over classical computation was suggested by Richard Feynman\cite{Feynman1982, Feynman1986a}.
He considered a universal quantum mechanical simulator, which is a controllable quantum system that can
be used to imitate other quantum systems, therefore being able to tackle problems that are intractable
on classical computers. Since then the motivation to use a quantum simulator as a powerful tool to address the most important
and difficult problems in multidisciplinary science has led to many theoretical proposals\cite{Lloyd1996, EdwardFarhi2000, Aspuru-Guzik2005, Trebst2006, Buluta2009, J.D.Biamonte2010}. Vast technological developments allowed for recent realizations of such devices in atoms\cite{Greiner2002, Lewenstein2007, Trotzky2009}, trapped ions\cite{Leibfried2002, Friedenauer2008, Gerritsma2010, Kim2010}, single photons\cite{Lu2009, Pachos2009, Lanyon2010, Lavoie2010} and NMR\cite{Somaroo1999, Du2010}. The quantum simulation of strongly correlated quantum systems (e.g.\ frustrated spin systems) is of special interest and would provide new results that cannot be otherwise classically simulated\cite{Verstraete2009}.

In order to manipulate and measure individual properties of microscopic quantum systems the complete control over all degrees of freedom for each particle is required. Typically, atoms in optical lattices\cite{Greiner2002} are used for realizing physical systems that can simulate various models in condensed-matter physics. The fact that the experimental addressability of single atoms in optical lattices remains very challenging\cite{Bakr2009, Bakr2010, Sherson2010} leads to the studies of bulk properties of the atomic ensemble ($\approx10^5$ atoms) instead of single particles. Therefore we utilize single photons in separate spatial modes and measurement-induced interactions as a quantum simulator, thus the particles are individually accessible. The tunable interaction between two entangled photon-pairs allows for the precise simulation of the ground state of a spin-1/2 tetramer. We obtain the ground-state energy and have direct access to the distribution of pair-wise quantum correlations as a function of the competing spin-spin interactions. We also observe the influence of monogamy\cite{Coffman2000a,Osborne2006} in this strongly correlated quantum system.

\section*{Analog quantum simulator}
The main challenge in the understanding of strongly correlated quantum systems is to calculate the energies
and ground state properties of many-body systems as this becomes exponentially difficult with increasing number of particles when using a classical computer. In contrast, quantum simulators use quantum systems to store and process data which allows them to polynomially mimic the evolution of the quantum system of interest. Typically, quantum simulations require methods to implement the Hamiltonian of the simulated system, and probe its ground state properties. In some cases, even if the ground state wave-function is known, obtaining all properties by using classical computers remains a challenging task. For example, for the one-dimensional antiferromagnetic Heisenberg model one can compute the spectrum and the wave-function by using the Bethe ansatz\cite{Bethe1931}, but it is not as easy to extract the correlations functions. This can be overcome by quantum simulators that are capable of generating and directly probing the quantum state of interest via controlling each quantum particle individually.

\begin{figure}
    \includegraphics[width=0.45\textwidth]{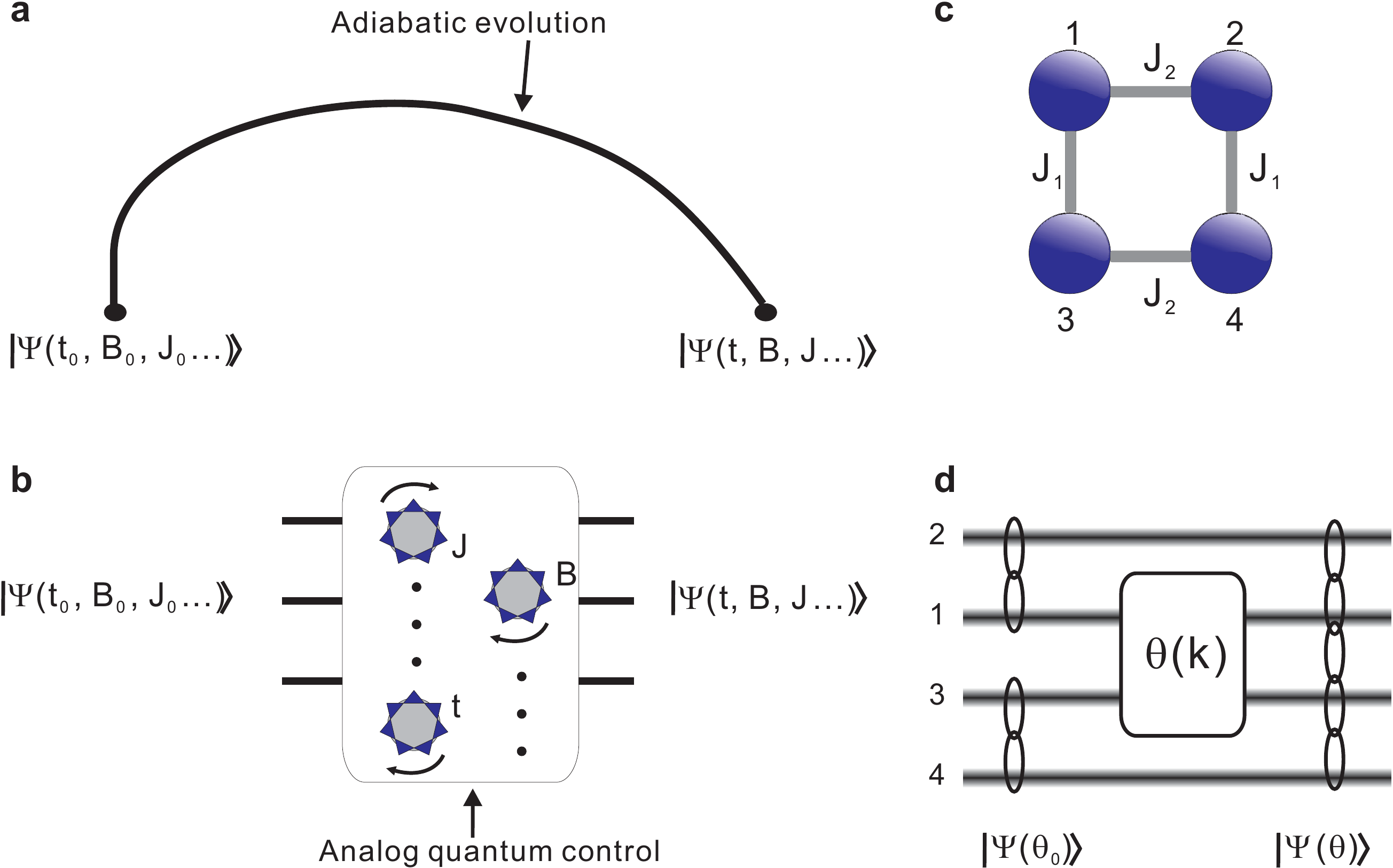}
    \caption{\label{figAQS}Mimicking an adiabatic quantum evolution with an analog quantum simulator. (\textbf{a}) Adiabatic quantum evolution. The system is prepared in an initial ground state $\ket{\psi(t_0,B_0,J_0,\dots)}$. Then the gradual change of the system parameters ($t$, $B$, $J$, etc.) causes an adiabatic evolution of the system to the final ground state of interest $\ket{\psi(t,B,J,\dots)}$. (\textbf{b}) Analog quantum simulation. The adiabatic evolution of the system to be simulated is mapped onto a controllable evolution of a quantum system. A set of tunable gates give access to the change of parameters. (\textbf{c}) Model used to study the valence-bond states. The nearest-neighbor Heisenberg-type interactions of strength $J_1$ and $J_2$ among four spin-1/2 particles are drawn as connecting bonds and form a spin-1/2 tetramer. All the properties of the tetramer depend only upon the coupling ratio $\kappa=J_2/J_1$. (\textbf{d}) Quantum simulation of a spin-1/2 tetramer using a photonic analog quantum simulator. The initial ground state, $\ket{\Psi(\theta_0)}$, is prepared by generating the photon-pairs 1 \& 2 and 3 \& 4 in two singlet states. Then the analog quantum simulation is performed utilizing the measurement-induced interaction, consisting of quantum interference and the detection of a four-photon coincidence after superimposing photons 1 \& 3 on a tunable beam splitter. Mapping the coupling ratio $\kappa$ on the beam splitter's splitting ratio $\tan^2\theta$, leads to the ground state of interest, $\ket{\Psi(\theta)}$.}
\end{figure}

Usually, the system being simulated is defined by its Hamiltonian $H(t,J,B,\dots)$ that is dependent on parameters such as time, $t$, interaction strength, $J$, external field, $B$, etc. In practise, one way of realizing a quantum simulator is
based on discrete gate operations and the phase estimation algorithm\cite{Aspuru-Guzik2005, Lanyon2010}, referred to as a digital quantum simulator\cite{Buluta2009}. An alternative approach utilizes the adiabatic theorem\cite{Born1928}, where an initial Hamiltonian, whose ground state is easy to prepare, can be adiabatically evolved to a final Hamiltonian with a nontrivial ground state of interest\cite{Lloyd1996, EdwardFarhi2000, J.D.Biamonte2010, Du2010}. An adiabatic quantum simulator can be built by engineering interactions among particles using tunable external parameters (e.g.\ an external magnetic field). The system will remain in its ground state if the system parameters change gradually enough.

Our experimental technique combines the advantages of both approaches by utilizing a tunable quantum gate without the necessity of either discretizing the quantum evolution or engineering the physical interactions for an adiabatic quantum simulation. Thus, we consider our simulator as an analog quantum computer\cite{Greiner2002, Friedenauer2008, Buluta2009, Bakr2009, Kim2010, Bakr2010, Sherson2010}, where the change of the quantum evolution can be obtained by a tunable quantum gate. The advantages of using analog quantum gate operations seem to be far-reaching as the number of gate operations can be less-demanding than for digital quantum simulators. Fig.\ \ref{figAQS} shows the concept of this analog simulator. In our experiment, we use the analog quantum simulator to direct probe the wave-function instead of the Hamiltonian. We take the advantage of single-particle addressability and the tunable interaction of our quantum simulator to prepare various ground states experimentally. Once the wave-function is prepared, it is feasible to measure all the correlation functions and study the dynamics of entanglement in this strongly correlated systems.

\section*{Simulation of a spin-1/2 tetramer}
Over the last 60 years the fundamental interest in studies of ground states of Heisenberg-type Hamiltonians has led to a few exact theorems, which may serve as guidelines for quantum simulators. Based on Marshall's theorem\cite{Marshall1955} and its extension\cite{Lieb1962} the absolute ground state has total spin zero ($S^2= 0$) for $N$ spins on a bipartite lattice with nearest-neighbor antiferromagnetic Heisenberg-type interactions.
This constraint leads to the fact that the ground state can be built as a superposition of pairs of spins forming singlets or equally so-called valence bonds. If all the spins are covered by valence bonds, which are maximally entangled states, then the ground state's total spin is zero and non-magnetic. This is established by valence bonds that are either static and localized or fluctuating as a superposition of different partitionings of spins. In general, the equally weighted superposition of all different localized valence-bond states corresponds to a quantum spin liquid, the so-called resonating valence-bond state\cite{Affleck1987, Balents2010}.

The smallest configuration for studying and simulating these phenomena on a two-dimensional square lattice is four spin-1/2 particles forming a tetramer. In the case of such a spin-1/2 tetramer the Heisenberg-type interactions lead to the creation of three possible dimer-covering configurations for the localized valence-bond states, $\ket{\Phi_=}\equiv\ket{\psi^{-}}_{12}\ket{\psi^{-}}_{34}$, $\ket{\Phi_\parallel}\equiv\ket{\psi^{-}}_{13}\ket{\psi^{-}}_{24}$ and $\ket{\Phi_\times}\equiv\ket{\psi^{-}}_{14}\ket{\psi^{-}}_{23}$, where $\ket{\psi^{-}}_{\textrm{ij}}$ is the singlet of particle i and j (Fig.\ \ref{figAQS}c). Since the total spin-zero subspace for this system is two-dimensional, these three dimer-covering states are not independent and $\ket{\Phi_\times}$ can be written as $\ket{\Phi_\times}=\ket{\Phi_\parallel}-\ket{\Phi_=}$ in the $\ket{\Phi_=}$/$\ket{\Phi_\parallel}$ basis. This state, $\ket{\Phi_\times}$, like any other equal superposition of these two dimer-covering states represents a resonating valence-bond state. Particularly interesting states are resonating valence-bond states\cite{Trebst2006, Mambrini2006}with s-wave pairing symmetry, $\ket{\Phi_\parallel}+\ket{\Phi_=}$ (up to normalization), and with the exotic d-wave pairing symmetry, $\ket{\Phi_\times}=\ket{\Phi_\parallel}-\ket{\Phi_=}$. The studies of these states are of high interest, because it was conjectured that a transition from an localized valence-bond configuration to the superposition of different valence-bond states might explain high-temperature superconductivity in cuprates\cite{Anderson1987}. A quantum simulator capable of preparing such arbitrary superpositions of dimer-covering states is thus sufficient for simulating any Heisenberg-type interactions of four spin-1/2 particles on a two-dimensional lattice. It is the particular strength of our optical quantum simulator that the simulated ground states can be restricted to the spin-zero singlet subspace by utilizing the quantum interference of photons at a tunable beam splitter.

For our quantum simulation we consider four spin-1/2 particles on a square lattice (tetramer) that interact via Heisenberg-type interactions. According to Marshall's theorem, the ground state is restricted to total spin-zero subspace for arbitrary antiferromagnetic interactions, independent of their type and strength. Here we model our spin tetramer with nearest-neighbor interactions of the strength $J_1$ and $J_2$ (Fig.\ \ref{figAQS}c) by the Hamiltonian
\begin{equation*}\label{Ham}
H=J_1\vec{S}_1\vec{S}_3+J_1\vec{S}_2\vec{S}_4+J_2\vec{S}_1\vec{S}_2+J_2\vec{S}_3\vec{S}_4,
\end{equation*}
where $\vec{S}_i$ is the Pauli spin operator for spin $i$. All the properties of the system depend only on the
coupling ratio $\kappa=J_2 / J_1$ therefore we re-normalize the
Hamiltonian to
\begin{equation}\label{Ham2}
H(\kappa)=H_{0}+\kappa H_{1},
\end{equation}
where $H(\kappa)=H /
J_{1}$ is the final Hamiltonian, $H_{0}=\vec{S}_1\vec{S}_3+\vec{S}_2\vec{S}_4$ the initial Hamiltonian and
$H_{1}=\vec{S}_1\vec{S}_2+\vec{S}_3\vec{S}_4$ the competing Hamiltonian of $H_{0}$.

Due to the simplicity of the quantum system we study here, we can calculate the expected ground state analytically, which allows us to verify the experimental results. The ground state of the Hamiltonian
given in Eq.~(\ref{Ham2}) is
\begin{equation*}\label{GS}
\ket{\Psi^{(0)}(\theta)}=\frac{1}{\sqrt{
n(\theta)}}(\cos2\theta\ket{\Phi_=}-\cos^2\theta\ket{\Phi_\parallel})
\end{equation*}
where $n(\theta)=\frac{1}{2}(\cos^4\theta+\cos^22\theta+\sin^4\theta)$ is
the normalization constantThe ground state energy of $\ket{\Psi^{(0)}(\theta)}$ is $E^{(0)}=-2(1+\kappa)-4
\sqrt{1-\kappa+\kappa^2}$.
The TDC's angle, $\theta$, takes the values from the interval of $0\leq
\theta \leq \frac{\pi}{2}$. Using our photonic quantum simulator, we can mimic the
adiabatic change of the Hamiltonian shown in Eq.~(\ref{Ham2}), where the
full range of the coupling ratio $-\infty \leq \kappa \leq +\infty$
is experimentally covered by tuning the angle of the TDC between
$\arctan\frac{1}{\sqrt{2}} \leq \theta \leq \frac{\pi}{2}$. For $\kappa=0$ ($\theta=\frac{\pi}{4}$), the ground state is $\ket{\Phi_\parallel}$, while for $\kappa=+\infty$ ($\theta=\frac{\pi}{2}$), the ground state changes to $\ket{\Phi_=}$. These two cases are dimer-covering states. Tuning the coupling ratio to $\kappa=-\infty$ results in the equally weighted superposition $\ket{\Phi_\times}+\ket{\Phi_\parallel}$, whereas $\kappa=1$ leads to the interesting resonating valence-bond state\cite{Anderson1987,Affleck1987,Balents2010} $\ket{\Phi_=}+\ket{\Phi_\parallel}$.

\begin{figure}
    \includegraphics[width=0.49\textwidth]{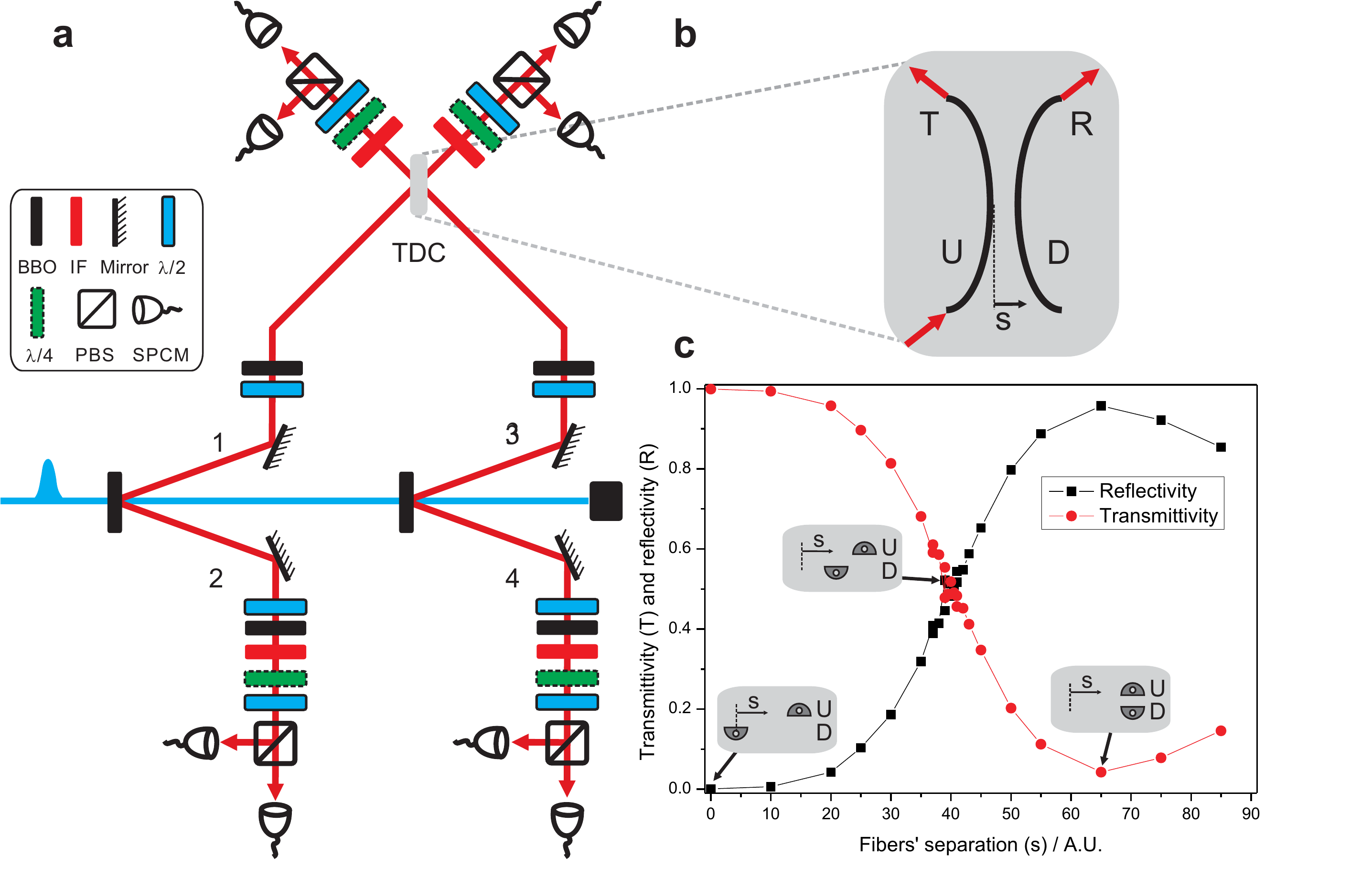}
    \caption{\label{figSETUP}Experimental setup.
    (\textbf{a}) Femtosecond laser pulses ($\approx140 \fs$, $76 \mhz$, $404 \nm$) penetrate two $\beta$-barium
    borate (BBO) crystals generating two pairs of photons in the spatial modes 1 \& 2 and 3 \& 4 (two-fold coincident count rate per pair $\approx 20 \khz$). The walk-off effects are compensated with a half-wave plate (HWP) followed by a BBO crystal in each mode.
    The photon's spectral and spatial distinguishability is erased with interference filters (IF, FWHM $=3 \nm$)
    and single-mode fibers. The polarization of each photon is analyzed by a combination of a quarter-wave plate (QWP), a HWP and a polarizing beam splitter (PBS). Single photons are detected by single-photon counting modules (SPCM). (\textbf{b}) Schematic diagram of the fiber-based tunable directional coupler (TDC).  The view from the top of the TDC illustrates the dependence of the coupling of the evanescent light upon the separation of the fibers. The coupling between these two fibers is controlled by adjusting the horizontal position of the D fiber. (\textbf{c}) Experimental calibration of TDC's transmissivity (red circles) and reflectivity (black circles) with respect to the position of the D fiber (s) is performed by using weak laser beams and SPCM. The separations of the fibers for 0\%, 50\% and 100\% transmissivity are shown in the insets. The experimental imperfections originate mostly from detector dark counts and the error bars, smaller than 0.5\% of the mean values, are based on a Poissonian distribution.}
\end{figure}

Here, we experimentally demonstrate an optical analog quantum simulator by producing two polarization
entangled photon pairs (see Fig.\ \ref{figSETUP}a), $\ket{\psi^{-}}_{12}=
\frac{1}{\sqrt{2}}(\ket{\mathrm{HV}}_{12}-\ket{\mathrm{VH}}_{12})$
and $\ket{\psi^{-}}_{34}=
\frac{1}{\sqrt{2}}(\ket{\mathrm{HV}}_{34}-\ket{\mathrm{VH}}_{34})$, in the spatial modes 1 \& 2 and 3 \& 4. $\ket{H}$ and $\ket{V}$ denote horizontal and vertical polarization states, respectively. The tunable interaction among these singlet states is achieved by the TDC, followed by a projective measurement of one photon in each of the four output modes. This tunability allows us to continuously change the measurement-induced interaction between photons 1 and 3. The TDC is an optical fiber device that transfers optical signals between fibers acting as a beam splitter with controllable splitting ratio. The control of the
splitting ratio is achieved by adjusting the relative positions of the fibers (Fig.\ \ref{figSETUP}b). The
transmissivity and reflectivity of this TDC go as $\cos^{2}\theta$
and $\sin^{2}\theta$, respectively, where $\theta$ parameterizes the separation of the fibers. We calibrate the TDC's
transmissivity and reflectivity such that the modulating visibilities (Michaelson visibility) are
above $95\%$ for both inputs, as required for high-precision quantum
control (Fig.\ \ref{figSETUP}c).

A successful detection of a four-fold coincidence event from
each spatial mode gives the four-photon state,
$\ket{\psi(\theta)}_{1234}  =
\frac{1}{\sqrt{n(\theta)}}[-\cos^{2}\theta(\ket{\mathrm{HHVV}}+\ket{\mathrm{VVHH}})
      +\sin^{2}\theta(\ket{\mathrm{HVVH}}+\ket{\mathrm{VHHV}})
      +\cos2\theta(\ket{\mathrm{HVHV}}+\ket{\mathrm{VHVH}})]$. The experimentally obtained density matrix, $\rho_{exp}$, is reconstructed
from a set of 1,296 local measurements using the maximum-likelihood
technique\cite{White1999, James2001}. For this, all combinations of
mutually unbiased basis sets for individual qubits, that is
$\ket{\mathrm{H/V}}$,
$\ket{\mathrm{+/-}}=\frac{1}{\sqrt{2}}(\ket{\mathrm{H}}\pm\ket{\mathrm{V}})$
and
$\ket{\mathrm{R/L}}=\frac{1}{\sqrt{2}}(\ket{\mathrm{H}}\pm\mathrm{i}\ket{\mathrm{V}})$, are measured.
The duration of each measurement for a given setting of the polarization analyzers and the TDC was 200~s and the average detected four-fold coincidence rate was 3~Hz. In total eight density
matrices for different settings of $\theta$ are reconstructed and
are summarized in the Supplementary Information. Uncertainties in quantities extracted from these density matrices are calculated using a 10 run Monte Carlo simulation of the whole state tomography analysis, with Poissonian noise added to each experimental data point in each run.

\begin{figure}
    \includegraphics[width=0.45\textwidth]{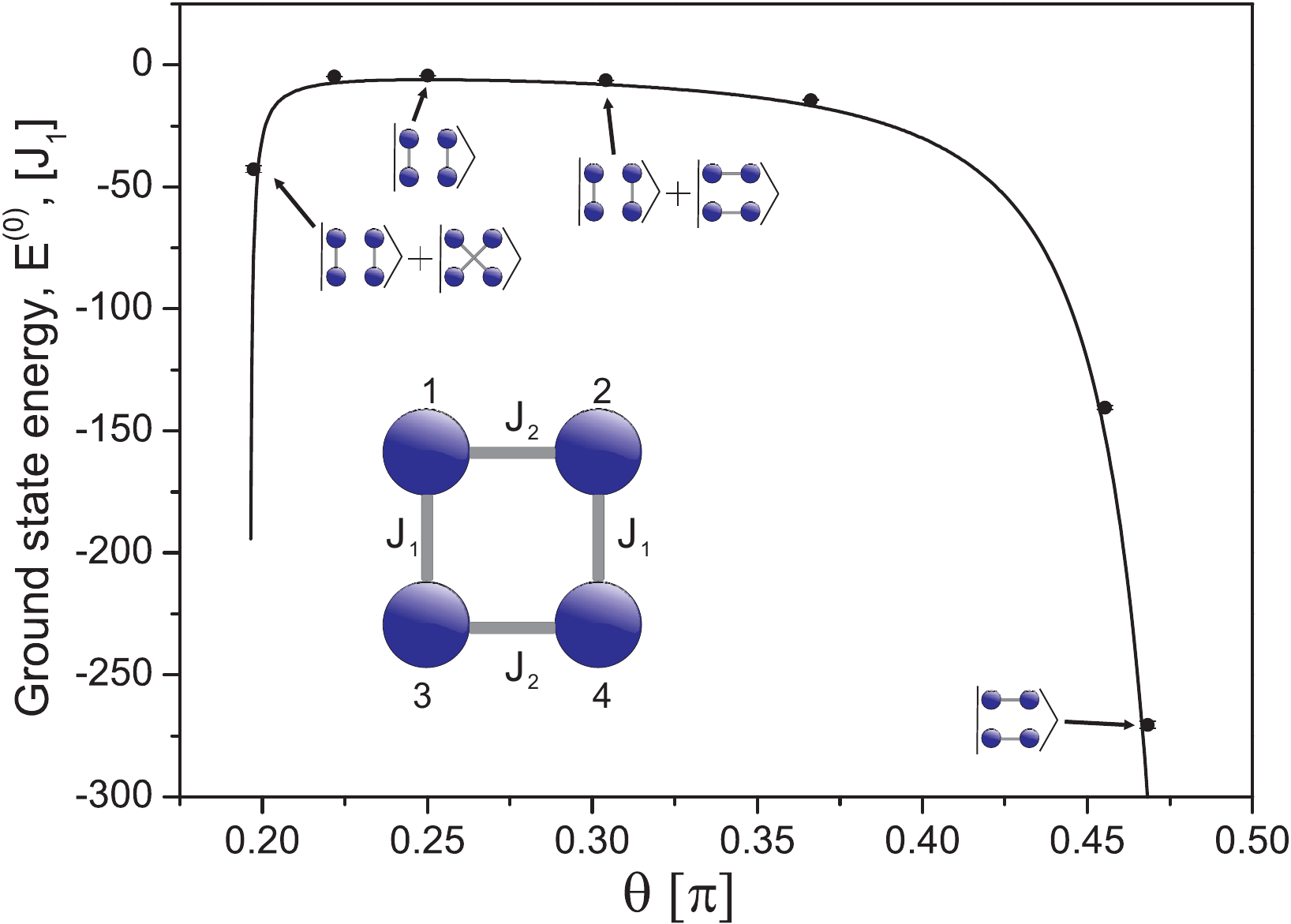}
    \caption{\label{figTE}Ground state energy of the spin-1/2 tetramer. By tuning $\theta$, where $\tan^2\theta=\kappa+\sqrt{\kappa^2-\kappa+1}$ represents the splitting ratio of the tunable directional coupler, we gradually change the ground state of the spin-1/2 tetramer. The full range of the coupling ratio $-\infty \leq \kappa=\frac{J_2}{J_1} \leq \infty$ is covered by tuning $\theta$ from $\arctan\frac{1}{\sqrt{2}}$ to $\frac{\pi}{2}$. We measure the ground state energy for seven different configurations. Of particular interest are the quantum states $\ket{\Phi_\parallel}+\ket{\Phi_\times}$, $\ket{\Phi_\parallel}$, $\ket{\Phi_\parallel}+\ket{\Phi_=}$ and $\ket{\Phi_=}$, shown explicitly. The black circles represent the experimental data and the solid line is parameter-free theoretical prediction. The error bars follow Poissonian statistics and are smaller than the data points.}
\end{figure}

The ground-state energy, $E^{(0)}$, is defined as the mean value of the Hamiltonian given in Eq.~(\ref{Ham2}). Since this Hamiltonian contains only pair-wise interactions, we measured the expectation value of the corresponding pair-wise correlations and obtained the data shown in Fig.~\ref{figTE}, which show good agreement with theoretical prediction.

\begin{figure}
    \includegraphics[width=0.49\textwidth]{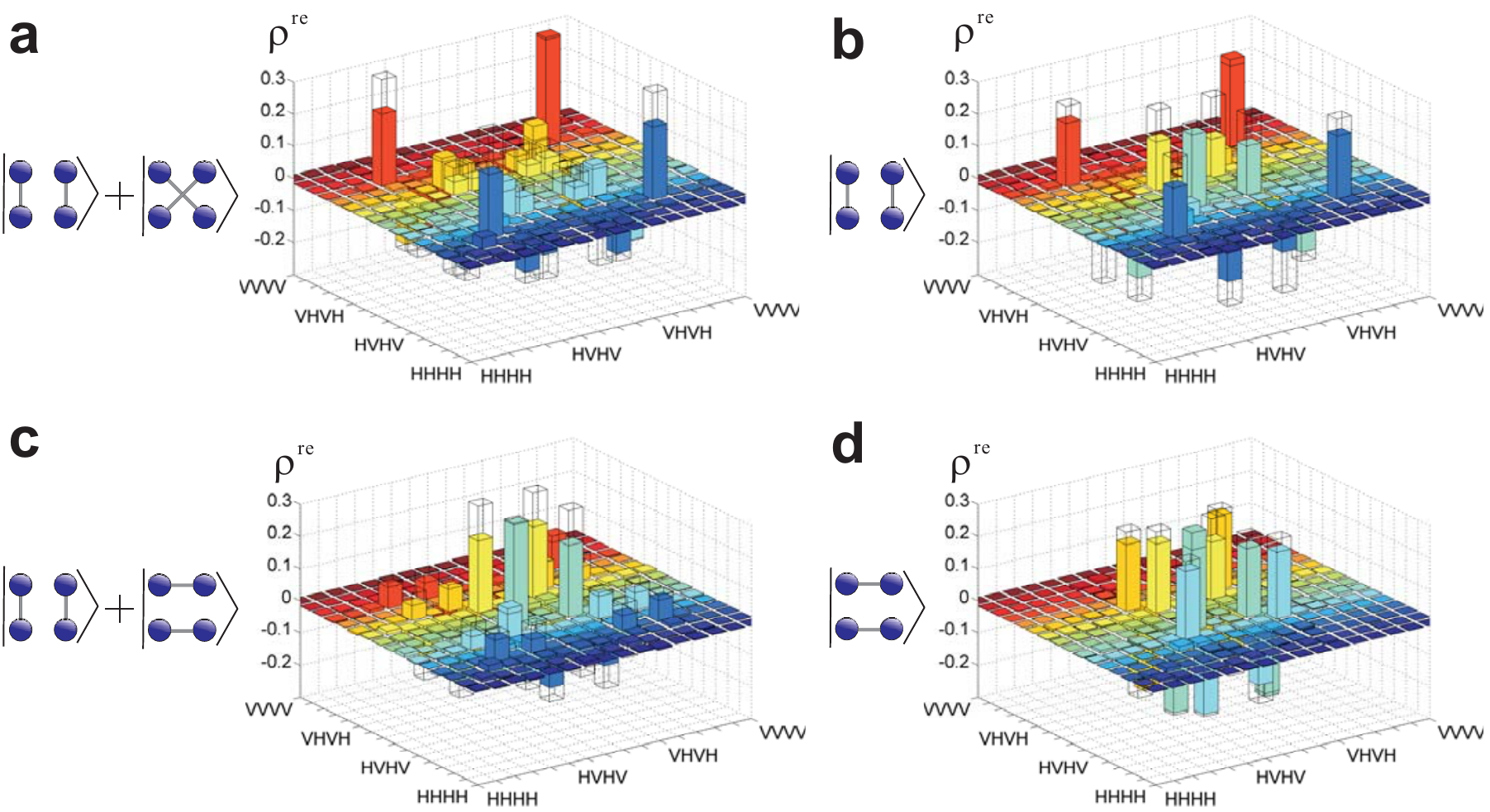}
    \caption{\label{figDM}Density matrices of various spin-1/2 tetramer configurations in the computational basis ($\ket{H}$/$\ket{V}$). Shown are the real parts of the density matrices for the cases of (\textbf{a}) equal superposition of dimer-covering states, (\textbf{b},\textbf{d}) dimer-covering states and (\textbf{c}) resonating valence-bond state. The imaginary parts are small and shown in the Supplementary Information. The wire grids indicate the expected values for the ideal case.  The density matrices are reconstructed from the experimental four-photon tomography data for the settings of (\textbf{a}) $\theta=0.197\pi$, (\textbf{b}) $\theta=0.25\pi$, (\textbf{c}) $\theta=0.304\pi$ and (\textbf{d}) $\theta=0.468\pi$. The fidelities, $F$, of the measured density matrix with the ideal state are (\textbf{a}) $F=0.745(4)$, (\textbf{b}) $F=0.712(4)$, (\textbf{c}) $F=0.746(6)$ and (\textbf{d}) $F=0.888(2)$. The uncertainties in fidelities extracted from these density matrices are calculated using a Monte Carlo routine and assumed Poissonian errors.}
\end{figure}

In Fig.\ref{figDM}, we show the experimentally obtained density matrices of the four valence-bond states, $\ket{\Phi_=}+\ket{\Phi_\times}$, $\ket{\Phi_\parallel}$, $\ket{\Phi_=}+\ket{\Phi_\parallel}$ and $\ket{\Phi_=}$, which correspond to the setting of $\theta=0.197\pi$, $\theta=0.25\pi$, $\theta=0.304\pi$ and $\theta=0.468\pi$. The state fidelity is defined as $F(\Psi, \rho)=\bra{\Psi}\rho\ket{\Psi}$, where $\ket{\Psi}$ is the target and $\rho$ is experimentally obtained quantum state. Due to the high quality of our quantum simulator, we obtain four-photon state fidelities that range from $F=0.712(4)$ to $F=0.888(2)$ (see Supplementary Information).

\section*{Quantum monogamy and complementarity}
Monogamy is one of the most fundamental properties of
quantum entanglement\cite{Coffman2000a,Osborne2006}. It restricts the shareability of quantum
correlations among parties and is of essential importance in many quantum information processing protocols, including quantum cryptography and entanglement distillation. Recent work showed that in the context of condensed-matter physics, monogamy gives rise to frustration effects in quantum system e.g. Heisenberg antiferromagnets. The ideal
ground state for an antiferromagnet would consist of singlets
between all interacting spins. But, due to the monogamy relation a particle can
only share one unit of entanglement (singlet) with its neighbors. Therefore, it will spread entanglement in an
optimal way with all its neighbors leading to a strongly
correlated ground state\cite{Osborne2006}.

\begin{figure}[ht!]
    \includegraphics[width=0.45\textwidth]{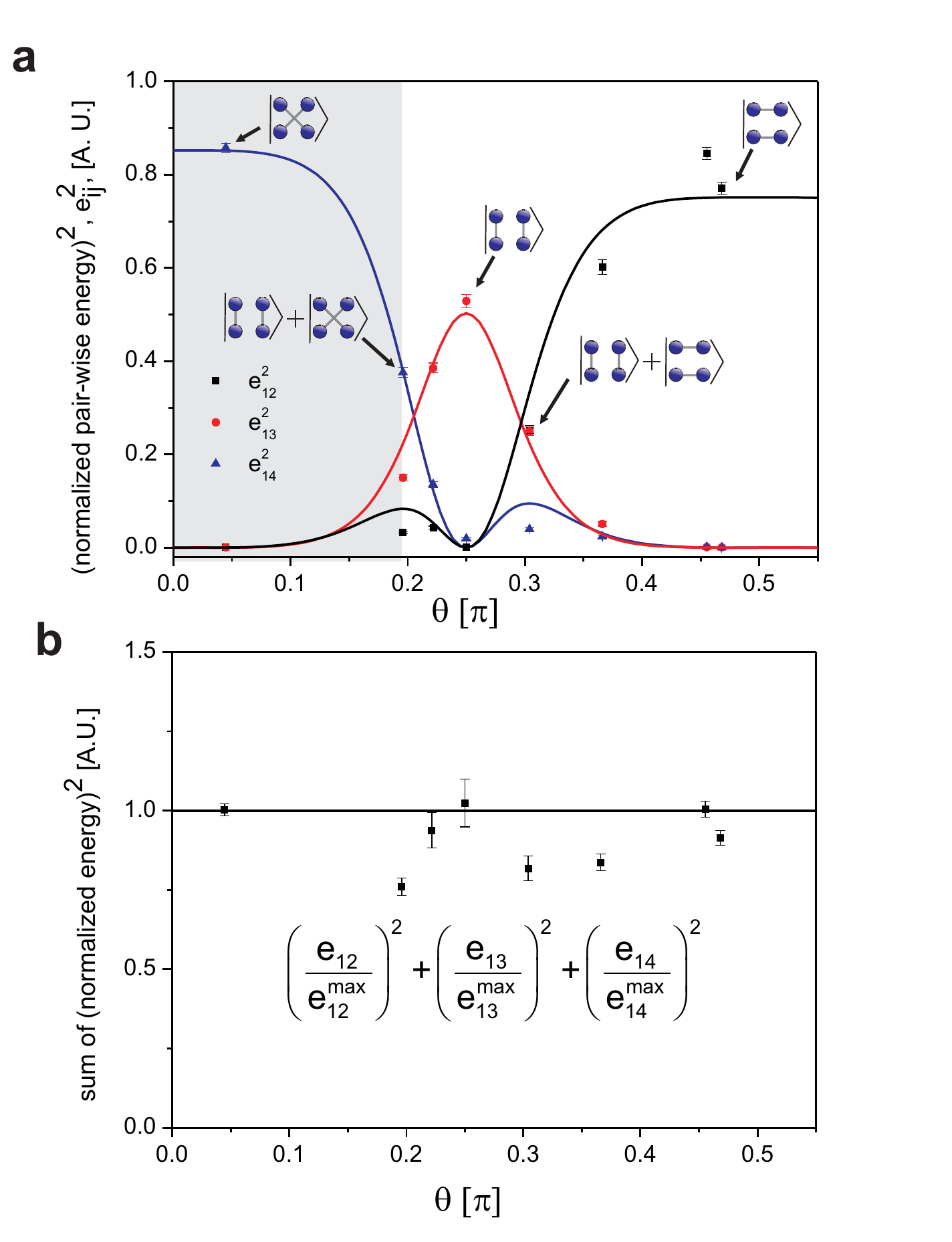}
    \caption{\label{figNE}Experimentally extracted pair-wise Heisenberg energies.
    (\textbf{a}) Experimental observation of quantum monogamy when comparing the pair-wise normalized Heisenberg energy, $e_{ij}$, it acts as a two-particle entanglement witness, for the spin pairs 1 \& 2 (black square), 1 \& 3 (red circle) and 1 \& 4 (blue triangle). The highlighted area corresponds to the full range of the coupling ratio $-\infty \leq \kappa=\frac{J_2}{J_1} \leq \infty$. For the case of $\kappa=0$ ($\theta=\frac{\pi}{4}$), the ground state of
    this spin-1/2 tetramer is $\ket{\Phi_\parallel}=\ket{\psi^{-}}_{13}\ket{\psi^{-}}_{24}$ and the amount of
    entanglement of the pair 1 \& 3 reaches its maximum while the pairs 1 \& 2 and 1 \& 4 are not entangled. Similarly, for the case of $\kappa=+\infty$ ($\theta=\frac{\pi}{2}$), the ground state is reduced to $\ket{\Phi_=}=\ket{\psi^{-}}_{12}\ket{\psi^{-}}_{34}$, where pair 1 \& 2 is now maximally entangled, and pairs 1 \& 3 and 1 \& 4 are disentangled. In the case of the resonating valence-bond state, entanglement distributions are equal between the pairs 1 \& 2 and 1 \& 3 (i.e.\ $e_{12}=e_{13}$). In other cases, entanglement is
    distributed according to the monogamy relation. (\textbf{b}) Experimental demonstration of the complementarity relation in a spin-1/2 tetramer. For each valence-bond configuration we measured pair-wise Heisenberg energies, $e_{ij}$, which are normalized by its maximal value, $e^{\mathrm{max}}_{ij}$. The sum of these renormalized energy values are in good agreement with the theoretical prediction (shown as line in the plot). The uncertainties represent standard deviations deduced from propagated Poissonian statistics.}
\end{figure}

To study the dynamics of pair-wise interactions, in which the monogamy of bipartite quantum entanglement distribution plays a crucial role, we characterize the distribution of the two-body energies and correlations between one spin with respect to the others with the normalized Heisenberg energy per unit of interaction, $e_{ij}$. It is defined as $e_{ij}=-\frac{1}{3}\mathrm{Tr}(\rho_{ij}\vec{S_i}\vec{S_j})$. Note that $\rho_{ij}$ is the density matrix of spins $i$ and $j$. The normalized Heisenberg energy per unit of interaction is also an entanglement witness\cite{Brukner2004,Amico2008} and reaches its maximum value of
$e_{ij}=1$ for the singlet state. The amount of entanglement can also be quantified by concurrence\cite{Wootters1998}, which is directly related to $e_{ij}$ with
$C(e_{ij})=\max\{0, -\frac{1}{2}+\frac{3}{2}e_{ij}\}$. For our four-spin system the dependencies of
the pair-wise energies with respect to the TDC's angle $\theta$ are given by
$e_{12}=-\frac{1}{n}(\sin^2\theta\cos2\theta)$,
$e_{13}=\frac{1}{n}(\sin^2\theta\cos^2\theta)$, and
$e_{14}=\frac{1}{n}(\cos^2\theta\cos2\theta)$.

\begin{figure}[t!]
    \includegraphics[width=0.49\textwidth]{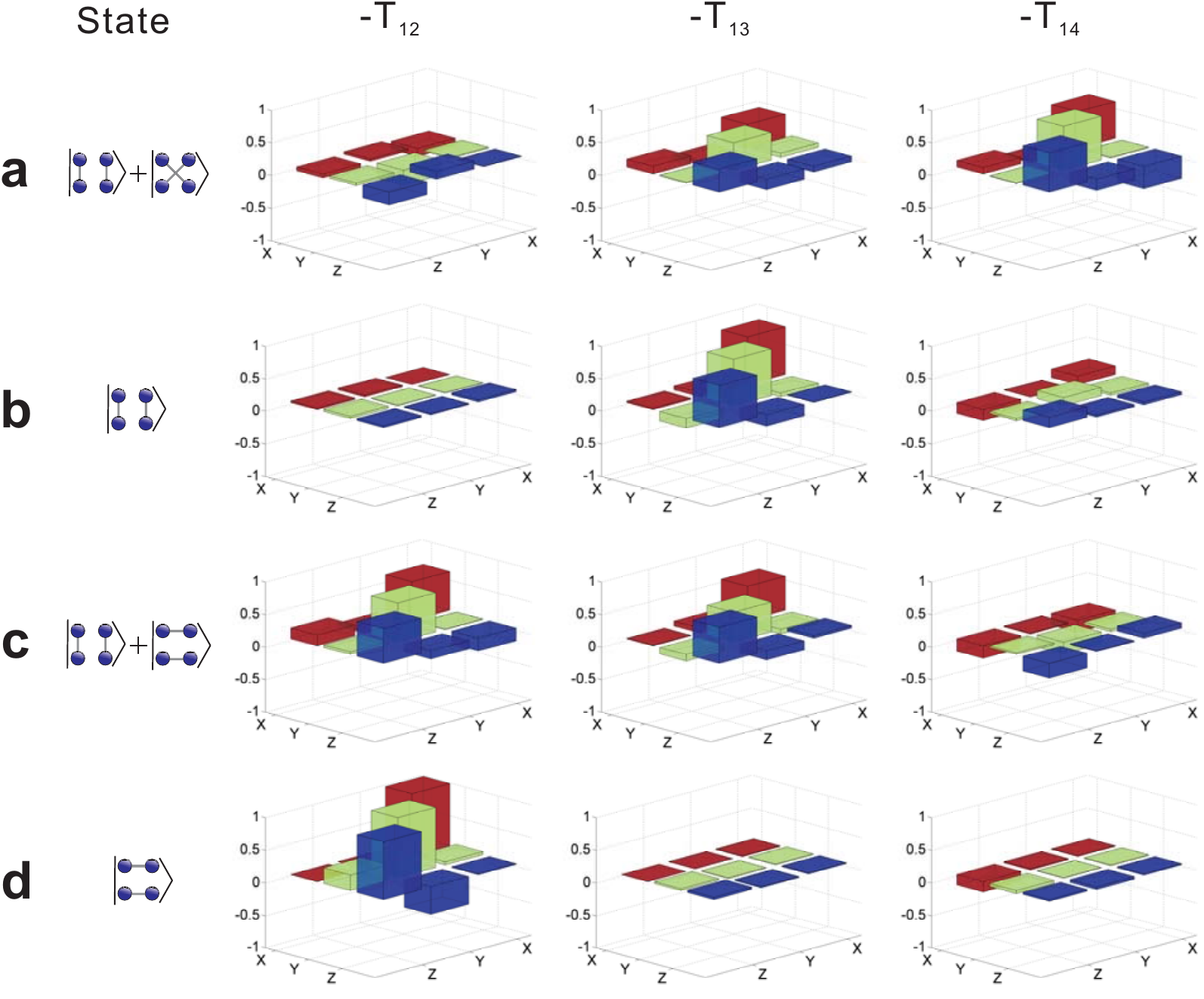}
    \caption{\label{figPC}Directly observed pair-wise correlation functions of various valence-bond states. The correlation tensors $T_{12}$ (photons 1 \& 2), $T_{13}$ (photons 1 \& 3) and $T_{14}$ (photons 1 \& 4) are obtained from correlation measurements directly in the bases $X=\sigma_{x}$, $Y=\sigma_{y}$ and $Z=\sigma_{z}$. For a convenient graphical representation, the negative values of the correlation tensors are shown. The structure of (\textbf{a}) the superposition state and (\textbf{c}) resonating valence-bond state show that the quantum correlations are equally distributed among two competing pairs. (\textbf{b}) and (\textbf{d}) belong to dimer-covering states, in which only one pair is maximally correlated in a singlet state.}
\end{figure}

Remarkably, monogamy is manifested in the constraint
of the energy distribution for the considered spin pair through a
complementarity relation\cite{Englert1996, Brukner2005}
\begin{equation}
\label{Eqcom}
e_{12}^2+e_{13}^2+e_{14}^2=1.
\end{equation}
This restricts the maximal amount of energy or
entanglement associated with correlated spin systems (see Fig.~\ref{figNE}a). For instance, in the experiment we obtain the normalized Heisenberg energy per unit interaction between photons 1 and 2 ($e_{12}$) with the correlation measurements in three mutually unbiased bases ($S_1^{(w)}\otimes S_2^{(w)}$, where $w=1,2,3$).

As shown in Fig.~\ref{figNE}a, the adiabatic change of the coupling between the four spins is simulated by tuning the angle of
the TDC, $\theta$, from $\arctan\frac{1}{\sqrt{2}}$ to $\frac{\pi}{2}$. This corresponds to the full
range of the coupling ratio $-\infty \leq \kappa=\frac{J_2}{J_1} \leq \infty$. In the ideal case, the maximum of $e_{ij}$ is unity which corresponds to a singlet
state shared by spins $i$ and $j$. However, imperfections in the generation of entangled photon pairs and
the two-photon interference on the TDC reduce the measured value of $e_{ij}$ by a constant factor,
independent of $\theta$. For the individual photon pairs we obtain the maximal Heisenberg energy of $e^{\mathrm{max}}_{12}=0.920(7)$, $e^{\mathrm{max}}_{13}=0.727(9)$, and $e^{\mathrm{max}}_{14}=0.926(5)$. In order to demonstrate the complementarity relation\cite{Durr1998} we re-normalized each energy $e_{ij}$ by its maximal value $e^{\mathrm{max}}_{ij}$ and obtain a good agreement with the theoretical prediction shown in Fig.\ref{figNE}b.

The advantage of the individual addressability for our particles in the ground state allows for the direct extraction of the pair-wise quantum correlations. The pair-wise quantum correlation is defined as:
\begin{widetext}
\begin{equation*}
T(S_i^{(w)}, S_j^{(v)})=\frac{C(S_i^{(w)}, S_j^{(v)})+C(S_i^{(w)\perp}, S_j^{(v)\perp})-C(S_i^{(w)\perp}, S_j^{(v)})-C(S_i^{(w)}, S_j^{(v)\perp})}{C(S_i^{(w)}, S_j^{(v)})+C(S_i^{(w)\perp}, S_j^{(v)\perp})+C(S_i^{(w)\perp}, S_j^{(v)})+C(S_i^{(w)}, S_j^{(v)\perp})},  \label{CF}
\end{equation*}
\end{widetext}
where $C(S_i^{(w)}, S_j^{(v)\perp})$ are the coincidence counts between pair $i$ and $j$ along the direction of $S^{(w)}$ and that perpendicular to $S^{(v)}$, respectively. In Fig.\ \ref{figPC}, the pair-wise correlation functions for the ground states, $\ket{\Phi_=}+\ket{\Phi_\times}$, $\ket{\Phi_\parallel}$, $\ket{\Phi_=}+\ket{\Phi_\parallel}$ and $\ket{\Phi_=}$, are shown. As expected from the monogamy relation, in the cases of dimer-covering states, one pair of the photons is maximally correlated, e.g.\ photons 1 and 3 in Fig.\ \ref{figPC}b, and photons 1 and 2 in Fig.\ \ref{figPC}c. In the cases of the equal superposition of two dimers (Fig.\ \ref{figPC}a) and the resonating valence-bond state (Fig.\ \ref{figPC}d), correlations are distributed among different pairs.

We demonstrate the feasibility of an all-optical analog quantum simulator by enabling quantum control of the measurement-induced interaction among photonic quantum states. Various ground states, including the resonating valence-bond states for four interacting spin-1/2 particles are generated and characterized by extracting the total energy and the pair-wise quantum correlations. The simulation of a spin-1/2 tetramer also proves that the pair-wise entanglement and energy distribution are restricted by the role of quantum monogamy. Our results provide promising insights for quantum simulations of small quantum systems, where individual addressability and control over all degrees of freedom on the single-particle level is required. This is of particular interest for quantum chemistry with small numbers of particles and might allow in the near future the simulation of aromatic systems and chemical reactions\cite{Kassal2008}. Although it was shown that efficient scalable quantum computing with single photons, linear-optical elements, and projective measurements is possible\cite{Knill2001}, the most important challenges will be (a) the realization of two- and more-qubit interactions with high fidelity, (b) generating systems with more qubits and (c) developing efficient methods of simulating other classes of complex Hamiltonians by using optical elements. Ideally, this and related work will open a new and promising avenue for the experimental simulation of various quantum systems.

\begin{acknowledgements}
The authors thank F. Verstraete, \v{C}. Brukner, W. Hofstetter, J. Kofler, T. Jennewein, R. Ursin, S. Zotter and S. Barz for discussions. We acknowledge support from the
European Commission, project QAP (No 015848), Q-ESSENCE (No 248095), an ERC senior grant (QIT4QAD), the Marie-Curie research training network EMALI, JTF, SFB-FOQUS and the doctoral programme CoQuS of the Austrian Science Foundation (FWF).
\end{acknowledgements}

\clearpage

\section*{Supplementary Information}
\subsection*{Quantum state tomography}
\begin{figure}[h!]
    \includegraphics[width=0.5\textwidth]{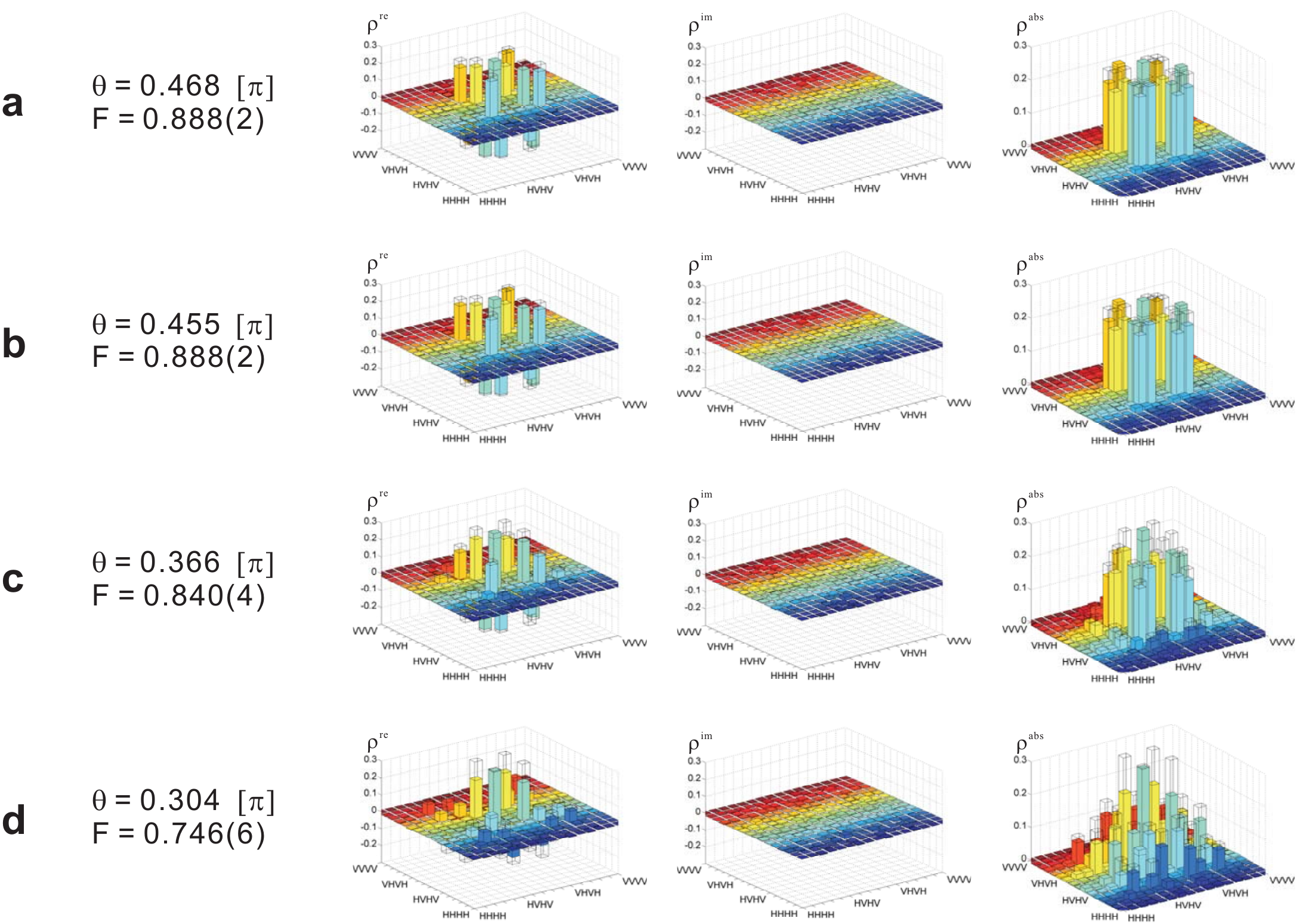}
    \caption{\label{smfig1}Density matrices of various spin-1/2 tetramer configurations in the computational basis ($\ket{H}$/$\ket{V}$). Shown are the real parts ($\rho^{re}$), imaginary part ($\rho^{im}$), and absolute values ($\rho^{abs}$) of the density matrices for the different settings of the splitting ratio of $\theta$ of the tunable direction coupler: (\textbf{a}), $\theta=0.468\pi$, (\textbf{b}), $\theta=0.455\pi$, (\textbf{c}), $\theta=0.366\pi$ and (\textbf{d}), $\theta=0.304\pi$. The wire grids indicate the expected values for the ideal case. The fidelities, $F$, of the measured density matrix with the ideal state are (\textbf{a}), $\textrm{F}=0.888(2)$, (\textbf{b}), $\textrm{F}=0.888(2)$, (\textbf{c}), $\textrm{F}=0.840(4)$ and (\textbf{d}), $\textrm{F}=0.746(6)$.}
\end{figure}

\begin{figure}[h!]
    \includegraphics[width=0.5\textwidth]{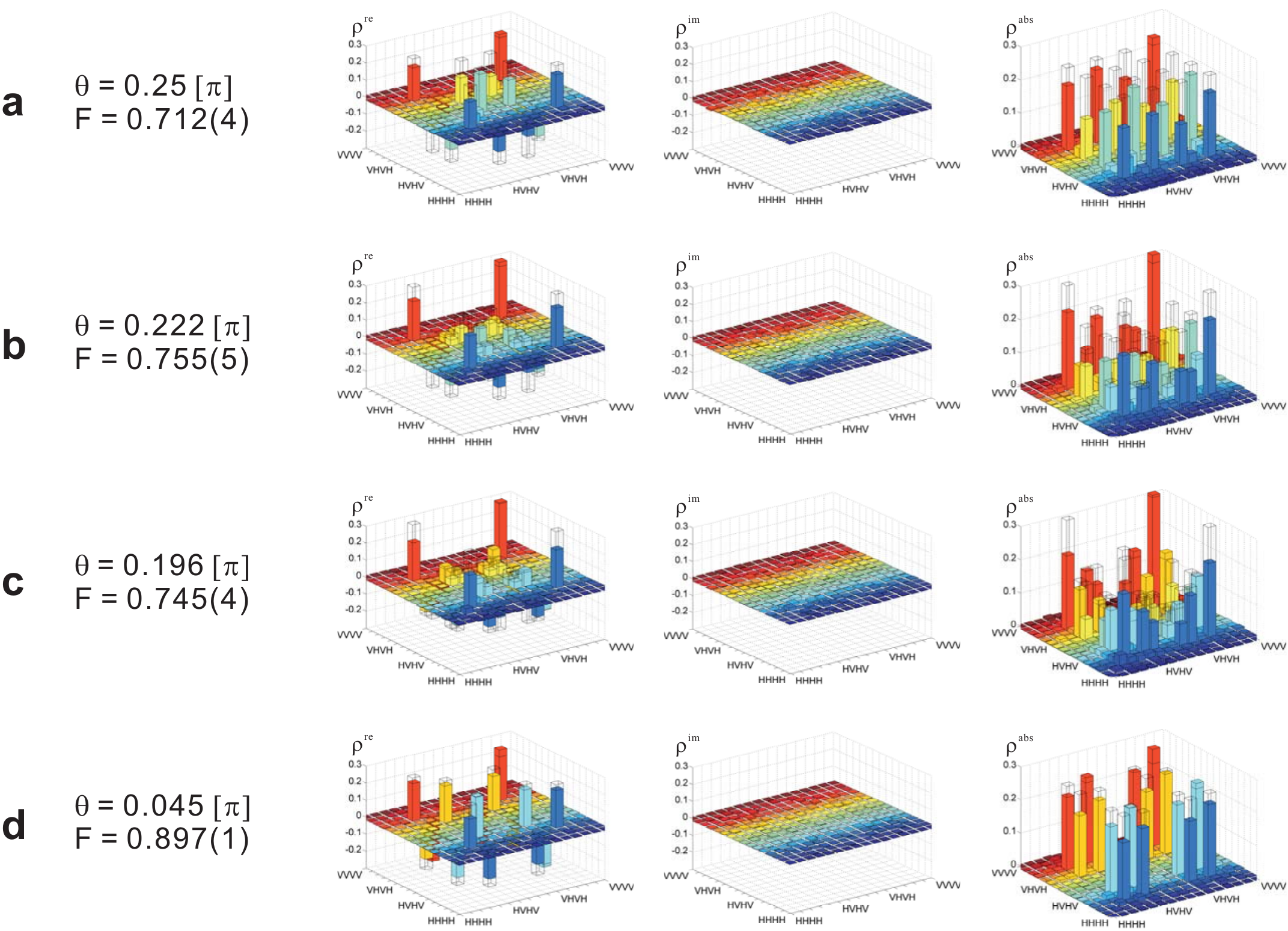}
    \caption{\label{smfig2}Density matrices of various spin-1/2 tetramer configurations in the computational basis ($\ket{H}$/$\ket{V}$). Shown are the real parts ($\rho^{re}$), imaginary part ($\rho^{im}$), and absolute values ($\rho^{abs}$) of the density matrices for the different settings of the splitting ratio of $\theta$ of the tunable direction coupler: (\textbf{a}), $\theta=0.25\pi$, (\textbf{b}), $\theta=0.222\pi$, (\textbf{c}), $\theta=0.197\pi$ and (\textbf{d}), $\theta=0.045\pi$. The wire grids indicate the expected values for the ideal case. The fidelity, $F$, of the measured density matrix with the ideal state are (\textbf{a}), $\textrm{F}=0.712(4)$, (\textbf{b}), $\textrm{F}=0.755(5)$, (\textbf{c}), $\textrm{F}=0.745(4)$ and (\textbf{d}), $\textrm{F}=0.897(1)$.}
\end{figure}

We perform quantum state tomography for the simulated ground states of a spin-1/2 tetramer. The studies of the eight different ground states lead to 10368 coincidence count measurements with a total 414724 four-fold coincidence counts. The reconstructed density matrices are plotted in Fig.\ \ref{smfig1} and \ref{smfig2}.

\end{document}